# Quantum key distribution with an efficient countermeasure against correlated intensity fluctuations in optical pulses

Ken-ichiro Yoshino[1], Mikio Fujiwara[2], Kensuke Nakata[3], Tatsuya Sumiya[4], Toshihiko Sasaki[4], Masahiro Takeoka[2], Masahide Sasaki[2], Akio Tajima[1], Masato Koashi[4] and Akihisa Tomita[3]

[1]NEC Corporation, [2]National Institute of Information and Communications Technology, [3]Hokkaido University, [4]The University of Tokyo

**Abstract**

**Quantum key distribution (QKD) allows two distant parties to share secret keys with the proven security even in the presence of an eavesdropper with unbounded computational power. Recently, GHz-clock decoy QKD systems have been realized by employing ultrafast optical communication devices. However, security loopholes of high-speed systems have not been fully explored yet. Here we point out a security loophole at the transmitter of the GHz-clock QKD, which is a common problem in high-speed QKD systems using practical band-width limited devices. We experimentally observe the inter-pulse intensity correlation and modulation-pattern dependent intensity deviation in a practical high-speed QKD system. Such correlation violates the assumption of most security theories. We also provide its countermeasure which does not require significant changes of hardware and can generate keys secure over 100 km fiber transmission. Our countermeasure is simple, effective and applicable to wide range of high-speed QKD systems, and thus paves the way to realize ultrafast and security-certified commercial QKD systems.**

## Introduction

Quantum key distribution (QKD) [1-3] allows two legitimate parties, Alice and Bob, to establish symmetric keys with the proven security even in the presence of an eavesdropper, Eve, who has unbounded computational power. Thanks to this unique feature, referred to as "information theoretic security", QKD, combined with Vernam's one-time pad cipher, enables the everlasting protection of confidentiality of data transmission, and hence must be an essential element to construct a long-term security system which cannot be realized only by cryptographic schemes based on computational security. Such a system has been exemplified in the literatures [4,5] as a long-term-secure storage network consisting of secret sharing, QKD and authentication schemes to deal with highly confidential data such as personal biomedical data, pharmaceutical, and genetic information. Because of growing interest in the confidentiality of those data, this storage network could be one of the killer applications of QKD.

Toward its practical realization, tremendous progress has been made during the past decades.

Metropolitan QKD networks have been successfully deployed [6-10] and is going to be a continental scale [11]. To provide information-theoretically secure keys to real applications securely and seamlessly, an efficient key management system and application program interfaces has been developed [12]. For the QKD device itself, high speed and stable operation is critical. By employing the ultrafast optical communication devices, high-speed QKD systems stably operated at GHz-clock frequency is realized in the installed-fiber networks [13-15].

Nevertheless, there remains an obstacle that makes the potential users hesitate to adopt this emerging technology; they would not innovate their existing secure-communication systems unless convinced that a QKD system at hand is really secure. In practice, like other cryptographic systems, a QKD system also has potential vulnerability due to mismatches between practical implementation and the theoretical model used for security proofs, which are referred to as side channels. For the QKD technology to be widely adopted, critical requirements are security certification, test-and-measurement method, security criteria for implementation, and countermeasures against the side channels. Moreover, those should be acceptable for non-experts. So far, receiver's security loopholes due to the side channels and countermeasures have been extensively studied for the existing QKD systems [16-19]. Also, the measurement device independent QKD protocol [20,21] can circumvent any receiver imperfections in principle. By contrast, researches on loopholes in transmitters have just begun in only a few aspects [22-26]. Loopholes in transmitters are directly linked to the mismatch of the state preparation between the ideal model and the implementation of QKD protocols. Therefore, rigorously quantitative evaluation of the imperfections in transmitters are essential to the security certification of QKD systems.

In this paper, we report a new security loophole, which may commonly exist in the transmitters of high-speed QKD systems, but has been overlooked so far despite its seriousness and generality. The current decoy-BB84 QKD systems generally rely on the matured ultrafast optical communication technology for high-speed operation, especially on signal modulation devices [13-15]. As shown below, the loophole in fact hides in the intensity modulator (IM) of such systems. Since practical modulators and electrical drivers are band-limited (which is common in optical communication as well), electrical signal distortion causes intensity correlation between the optical pulses as well as intensity fluctuation of individual pulses, where the former is particularly critical for the security since current security analysis usually assumes independent and identically distributed (IID) pulses. Such an inter-pulse intensity correlation occurs inevitably, and would provide additional information to an eavesdropper (Eve) to distinguish decoy state pulses from signal pulses. In other words, the QKD system without the countermeasure against intensity correlation in optical pulse train can be no longer guaranteed secure, and such a defective QKD system may cause a disaster for secure communications.

Against such a serious loophole, we develop its countermeasure which does not require new hardware. Although there are previous works [27-29] extending the coverage of the security proofs to accommodate the non-IID cases, a better performance of the QKD system will be achieved by developing more preemptive methods to circumvent correlations and fluctuations, based on the understanding on the real GHz-clocked QKD system [12] characteristics. We experimentally observe this modulation-pattern dependent intensity deviation and provide an efficient countermeasure. Our countermeasure consists of three post-processing operations: pattern sifting (PS), alternate key distillation (AKD), and intensity sifting (IS), which effectively recover the IID assumption and work for finite key length. Finally, we estimate the secure key rate and confirm achievability of the transmitter-loophole-closed secure key generation by high-speed decoy QKD system over 100km.

**Optical intensity deviation with inter-pulse correlation**

Figure 1 shows a conceptual view of our QKD transmitter working at 1.24-GHz clock rate. The first intensity modulator (IM) controls the intensity of the 50ps-width laser pulse for the three-state decoy protocol [30,31] and the following devices are for the time-bin BB84 signal encoding by an asymmetric Mach-Zehnder interferometer (AMZI), a modulator for encoding, and a variable optical attenuator (VOA) to attenuate the pulse energy to the single-photon level. The decoy IM is a dual-electrode Lithium Niobate (LN) modulator of 10 GHz bandwidth, driven by an electrical circuit designed for 10 Gbps digital optical communication. Relative input timing of optical pulses and modulation signals to the IM is controlled by fiber length connected in front of the IM with the accuracy of 50 ps (corresponding to fiber length of 1 cm).

The three-state decoy pulses are generated as follows. Two phase shift parameters $\varphi_i$ ($i = 1,2$) in the waveguides determine the output intensity as $I_{out} = \cos^2[(\varphi_1 - \varphi_2)/2] I_{in}$: $\{\varphi_1, \varphi_2\} = \{0,0\}$ for "signal" (S), $\{\varphi_1, \varphi_2\} = \{\pi, 0\}$ for "vacuum" (V), and $\{\varphi_1, \varphi_2\} = \{\pi, \phi\}$ for "decoy" (D) states, where $\phi$ is determined by the designed decoy intensity. The phase shifts $\pi$ and $\phi$ are generated by electrical voltage pulses with the heights of $V_\pi$ and $V_\phi$, respectively. We assign voltages of $V_\pi$ for "Hi" and 0 for "Lo" as the driving signals to one electrode (signal 1), whereas $V_\phi$ for "Hi" and 0 for "Lo" as the one to the other electrode (signal 2), respectively. Using these assignments, S state can be generated by {signal 1, signal 2} = {Lo, Lo}, and V state by {Hi, Lo}, so that transmittance of the IM takes the maximum and minimum values at these applied voltages. On the other hand, D state can be produced by either {Lo, Hi} or {Hi, Hi}. In our case, {Hi, Hi} is used because required value of $V_\phi$ is smaller than {Lo, Hi} for typical intensities where D state intensity is less than half of S state. Generally, an IM needs to be operated with mark rate of 50% to suppress the charge drift in the LN modulator during long-term operation. If the mark rates of modulation signals are biased, spontaneous polarization in the LN crystal is gradually enhanced, and it results in the waveform distortion of optical pulses. Our IM is operated with complementary mode, in which binary electrical signals Hi (Lo) in the first half of the pulse period is inverted to Lo (Hi) in the second half, automatically achieving mark rate of 50%.

If the modulation worked perfectly for the randomly chosen S, D, and V states, the intensity of the optical pulses should be determined independently without fluctuation. However, in real high speed systems, the electrical waveform distortion and the timing jitter of the optical pulses will cause unwanted intensity change depending on the state of the preceded pulse, *i.e.,* the intensity becomes correlated. We call this phenomenon as "pattern effect". A simple explanation of the pattern effect is as follows; an ideal drive circuit with a flat frequency response provides rectangular waveforms to the IM. The pulse amplitudes are independent of the previous modulation signal, as shown by the waveforms in Fig. 1 (a). However, the frequency response of real drive circuits is not uniform; it may show resonant peaks, and reduction in high frequency signals. Such imperfect frequency response distorts the waveform as shown in Fig. 1 (b). The electrical signal amplitude may differ according to the previous modulation patterns. This phenomenon results in correlated intensity deviation of modulated optical pulses.

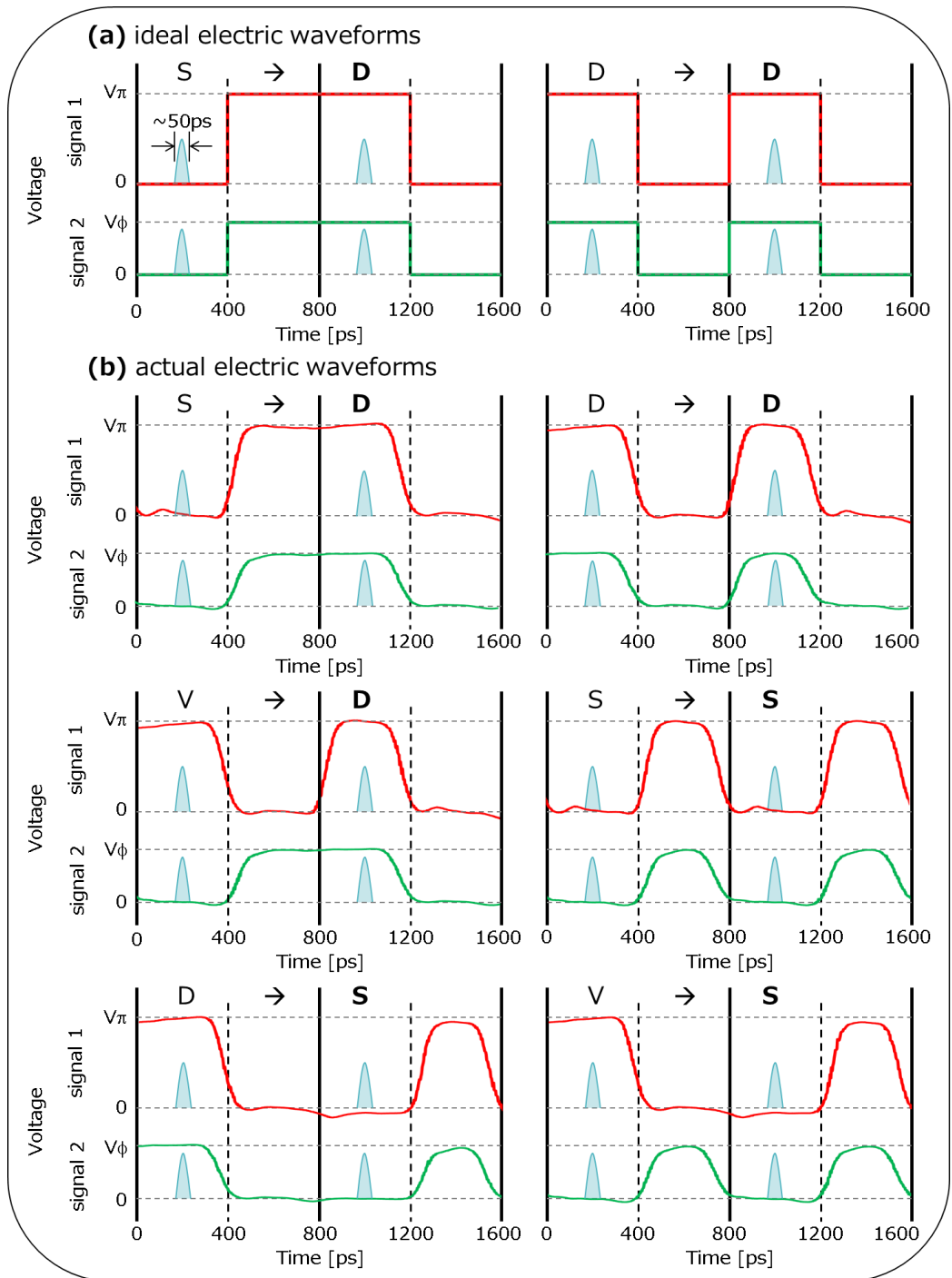

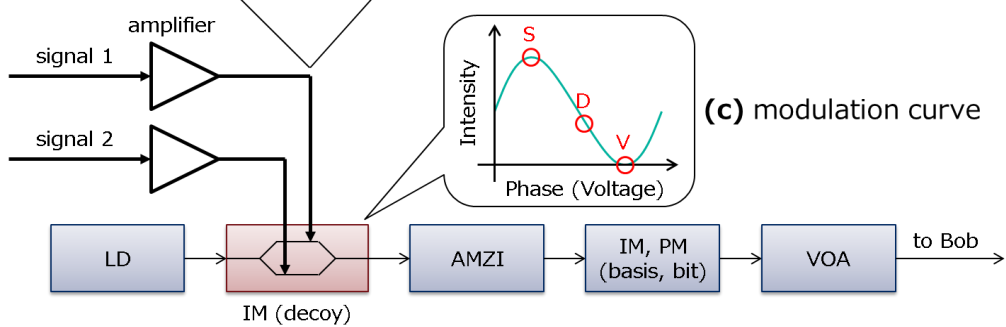

**Figure 1: Conceptual view of a transmitter (Alice) in typical decoy-BB84 QKD system using time-bin coding.** LD: laser diode, IM: intensity modulator, AMZI: asymmetric Mach-Zehnder interferometer, PM: phase modulator, VOA: variable optical attenuator.
(a) Ideal waveforms to the IM encoding signal (S) and decoy (D) state with complementary modulation. Pulse-shaped figures represent input timing of optical pulses. Pulse period (0.8 ns) is defined by the two solid lines. (b) Actual distorted waveforms from the 10-GHz bandwidth circuit. Optical pulses for decoy state with the preceding pulse D experience smaller phase shift than that with S. (c) conceptual image of operation points of intensity modulation.

We measured the pattern effects by picking the optical pulses from the output of the IM. The optical pulses are measured by a high speed photo-receiver with 9.3 GHz-bandwidth and subsequently recorded in an oscilloscope with 8 GHz-bandwidth. We defined the pulse intensity as the area of the time profile of the measured signal in one period containing a pulse peak. This measurement can evaluate the energy of each optical pulse. We measured 100,000 pulses per single pulse pattern for statistical analysis. In this experiment, we evaluated the pulse intensity before strong attenuation by a VOA, assuming that the intensity fluctuation of the optical pulses linearly reflects the mean photon number fluctuation in the quantum pulses through heavy attenuation.

In the following, we will show that measurements on only six pulse patterns among countless patterns of previous pulses are enough to characterize the pattern effect in our QKD equipment. The main cause of the pattern effect is the limited frequency response of the driving circuit, as explained before. The electric waveforms in Fig.1 (b) shows that modulation signals for the first pulses arrive at setting levels (0 or $V_{(\pi,\phi)}$) in 800 ps. It implies that the first half of the modulation signals has little effect on the second half. In other words, almost all the influence of pattern effect are limited to the adjacent pulses. Therefore we consider only the adjacent pulse states. Note that even if the complementary operation is used, pattern effects between adjacent pulses appear when the bandwidths of the devices are not enough because the voltage of the second half of previous modulation is different depending on its own signal (S, D or V). We ignore the intensity fluctuation of V state, since its effect on photon detection is smaller than that of stray light and dark counts. Then, we only need to measure the intensities of S pulses and D pulses with three types of predecessors: S, D, and V pulses. The six patterns are abbreviated to S→S, D→S, V→S, S→D, D→D, and V→D, where the intensities of the second pulses are to be measured.

Table 1 lists the averaged pulse intensities for the six patterns. While pattern effects were small (0.6 %-2.1 %) on the S pulses, large deviation about 20 % was observed on the D pulses. The deviation exceeded the normalized standard deviation of the intensity fluctuation around 7-9 %.

The different behavior of the D pulses comes from the operating point of the decoy pulses. At this point, the output intensity is sensitive to the applied voltage fluctuation as depicted in Fig. 1 (c). In contrast, those of the vacuum and signal pulses are set to the extreme of the input-output characteristics of the modulator, so that the output intensities are insensitive to the applied voltage.

One may consider the band-limitation of the measurement devices created "fake" pattern effects. If so, the pattern effect should have also appeared in S-state. However, observation showed that the pattern effect occurred only in D-state. Therefore, we concluded that the pattern effect originated from the intensity modulator.

When such a correlated intensity deviation is apparent, IID property of the pulse sequence is not approved. Therefore, conventional security analyses can be no longer applied directly. This issue appears to varying degree as long as the operating point is set on the steep slope of the modulation curve, shown as point (D) in Fig. 1 (C). We propose a simple and effective solution in the next section.

**Table 1: Measured intensity of signal pulses and decoy pulses for three types of predecessors.**
S: signal pulse, D: decoy pulse, V: vacuum state. Deviation of the intensity is given by that from the reference patterns (S→S and S→D).

| pattern | average intensity of second pulse (S) | deviation from S→S | normalized standard deviation |
|---|---|---|---|
| S → S | 1.000 +/- 0.032 (a.u) | ----- | 0.032 ($\tilde{\sigma}_S$) |
| D → S | 1.021 +/- 0.033 (a.u) | + 2.1% | 0.032 |
| V → S | 1.006 +/- 0.034 (a.u) | + 0.6% | 0.034 |

| pattern | average intensity of second pulse (D) | deviation from S→D | normalized standard deviation |
|---|---|---|---|
| S → D | 0.421 +/- 0.030 (a.u) | ----- | 0.070 ($\tilde{\sigma}_D$) |
| D → D | 0.344 +/- 0.031 (a.u) | - 18.2% | 0.090 |
| V → D | 0.331 +/- 0.030 (a.u) | - 21.4% | 0.091 |

## Countermeasure to the pattern-effect loophole: pattern sifting and alternate key distillation

Here we provide a software-based countermeasure against the pattern effects called "pattern sifting (PS)" and "alternate key distillation (AKD)". We ignore the minor deviations observed in D->S and V->S, which are smaller than standard deviations, and assume that the intensity of an S pulse is independent of its predecessor. PS discards particular modulation patterns in the key distillation process. The sifting rule on a pulse should be independent of its nominal intensity S, D, or V. Otherwise sifting itself may offer information on the intensity to Eve. In other words, we should decide whether we discard the focused pulse using the knowledge of other pulses. As mentioned in the previous section, since we need to consider the correlation only between the adjacent pulses in our QKD transmitter using the complementary modulation, the sifting rule should depend on the state of the adjacent pulses. The effect of the predecessor pulse can be avoided by fixing its nominal pulse intensity. The most efficient choice is to discard the pulse whose predecessor pulse is in D or V states, while sifting out the ones preceded by S pulses. The correlation with successor pulse can be disregarded by discarding the pulse whose successor is in D state, because the D state intensity is affected by the focused pulse intensity. The rule is summarized as follows:

(A) Discard the pulse, if its predecessor is in D or V state.

(B) Discard the pulse, if its successor is in D state.

Pulses are discarded depending on the state of predecessor and successor, not on the state of target pulse itself. Therefore, proportion of S, D and V is unchanged. As a result of the PS, the statistics of the sifted even-indexed pulses becomes IID conditionally on the variables for the odd-indexed pulses, and the same goes for the sifted odd-indexed pulses.

After the PS process, we divide sifted keys into odd-indexed events and even-indexed events according to the emission time stamps, and execute key distillation for each bit sequence. We refer to this process as AKD. Although there are no correlations among the even-indexed pulses conditionally on the variables for the odd-indexed pulses, there still remains a possibility of correlations between the even-indexed pulses and the odd-indexed ones. This makes it rather nontrivial whether both of the odd and even keys from the AKD are simultaneously secure. We solved this issue by dividing known security proofs of a decoy-state BB84 protocol into two statements, one for estimation of photon number statistics through the use of decoy states and the other for security of a BB84 protocol with an imperfect source. We then found that each statement allows composition of the even and odd parts, namely, that a statement for the even-indexed part and one for the odd-indexed part together imply a similar statement for the whole, regardless of correlations. The detail is given in the Methods section.

Figure 2 summarizes sifting rules in PS and pulse selection rules in AKD. Upper table regarding PS shows the probabilities for the pulse patterns using typical values of selection probabilities of signal, decoy, and vacuum states $p_S$=14/16, $p_D$=1/16, and $p_V$=1/16. After PS, $p_S(1-p_D)$ of the pulses will contribute to the key distillation, where the first factor comes from PS (A) and the second from PS (B). This fraction is 0.82 with the typical values, so that we can use most of the pulses for key distillation.

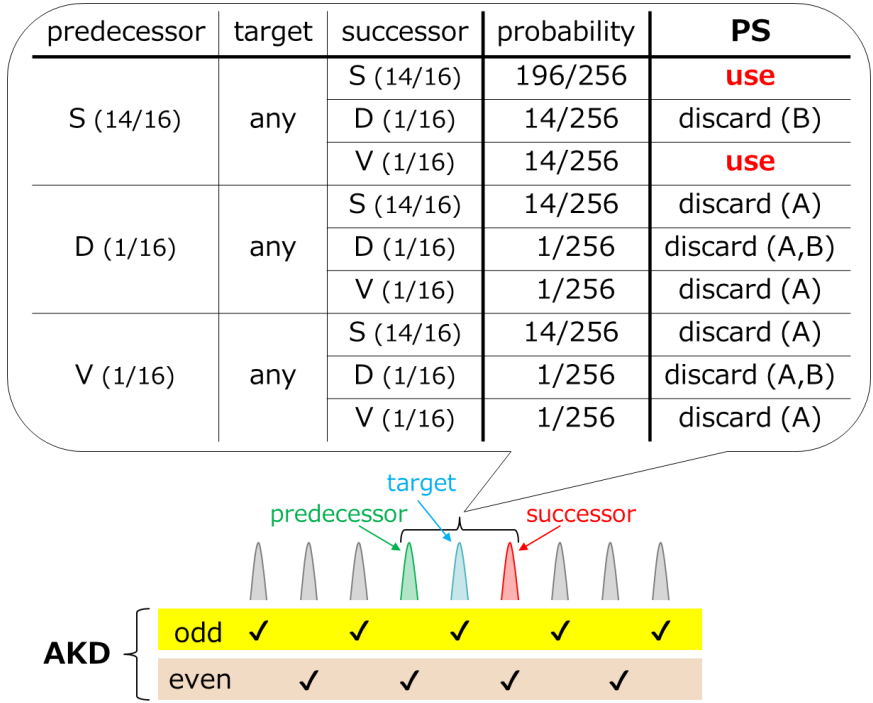

**Figure 2: Summary of sifting rules in "pattern sifting (PS)" and pulse selection rules in "alternate key distillation (AKD)".** The numbers after S, D, and V show the typical values of the selection probability.

The pattern effect can also be avoided by following naive protocol. If Alice sends a pulse with a fixed intensity before the pulse used for key distillation, no pattern effects would be observed. For example, Alice always selects S-state for odd number pulses, then the intensity of even number pulses are immune from the pattern effects. However, in this protocol, Alice and Bob should discard the odd number pulse outcomes, because Eve may also know the intensity of the odd number pulses and improve her measurement for successful eavesdropping. Therefore, the final key rate in the naive protocol is decreased to half of the original protocol.

Furthermore, one may think that faster devices can avoid the pattern effects. However, it is not clear how much bandwidth is needed for individual QKD systems, and it takes a very high cost, which is an obstruction against the widespread use of QKD systems. Our software-based PS and AKD enable to generate secure key using an existing QKD system without hardware replacements.

**Finite-length analysis with intensity sifting**

As long as the actual correlated pulse sequence is stationary, PS and AKD enable us to treat sifted key as if it was generated from an IID pulse sequence. Nevertheless, we have to consider the residual random intensity fluctuation, which would be brought by thermal noise or timing jitter of optical pulse and modulation electrical signals. Output power of the LD also would fluctuate. One way to establish a secure key in the presence of such a fluctuation is to apply "intensity sifting (IS)" to bound maximum and minimum of pulse intensities. In IS, we omit pulses whose intensities exceed the bound from key distillation process. This can be implemented with a pulse intensity monitor before attenuation by VOA in the transmitter and screening of the events for key distillation.

We evaluated standard deviations in the second pulse and normalized them using the average intensities of S and D as shown in Table1. We referred standard deviations of the second pulse in the case of S→S and S→D patterns as $\sigma_S$ and $\sigma_D$, and normalized standard deviations as $\tilde{\sigma}_S$ and $\tilde{\sigma}_D$, respectively. The obtained values of $\tilde{\sigma}_S$ and $\tilde{\sigma}_D$ were 3.2% and 7.0%. The fluctuation of the decoy intensity $\tilde{\sigma}_D$ was larger than that of the signal $\tilde{\sigma}_S$, because of the steep slope of the modulation curve as depicted in Fig.1 (c).

We extend a finite-key analysis [32] to consider the intensity fluctuations in signal, decoy and vacuum pulses. In Ref [32], authors provided concise finite-key security bounds which is based on the asymmetric decoy-state analysis proposed in Ref [30]. The IS procedure assures that the mean photon number of each pulse stays within the range $[\mu_a^L, \mu_a^U]$ ($a = S, D, V$). We rederived the key length formula of Ref [32], which yields smallest final key rate by considering the range of the mean photon number. Details of the reformulation are described in Supplementary Information.

Combination of PS, AKD and IS enables secure key generation even if the QKD equipment has correlated intensity deviation due to the pattern effect and random intensity distribution due to the thermal noise or timing jitter. We estimate the secure key rate of our GHz-clock QKD system. We assume that the intensity fluctuation obeys Gaussian distribution for S and D. We set the intensity range as $[\mu_a - t\sigma_a, \mu_a + t\sigma_a]$ with a common factor $t$ multiplied to standard deviation $\sigma_a$ for $a$ = S, D. We model the intensity fluctuation of the vacuum signal V by a half-Gaussian distribution $c \exp[-\mu^2/(2\sigma^2)]$ ($c$ is a normalization constant, and $\mu \geq 0$) and assume that its magnitude is similar to that of S, namely, $\sigma = \sigma_S$. The intensity range in IS is set to $[0, t\sigma]$. Note that such Gaussian assumptions are not necessary in practice, since we can calibrate the probabilities of passing the IS and calculate effective probabilities $p_S$, $p_D$, and $p_V$ accordingly.

We evaluate secure key rate with the three state decoy protocol with the nominal intensities $\mu_S$=0.5, $\mu_D$=0.2. We assume that Alice selects Y-basis (Z-basis) with the probability of $P_{Ya}$=0.25 ($P_{Za}$=0.75), and Bob adopts passive basis choice by a fiber splitter to feed photon pulses to a single photon detector (InGaAs/InP APDs) with the $P_{xb}$ for each basis ($x$=Y, Z). We set the probabilities to $P_{Yb}$=0.25 and $P_{Zb}$=0.75. The detector performances are assumed as follows: detection efficiency of $\eta_{det}$=0.1, dark count probability $P_{dc}$ of $10^{-6}$ and after pulse probability $P_{ap}$ of $10^{-2}$. Transmittance of the optical devices in Bob $\eta_{Bob}$ is assumed to be 0.25. We assume that the fiber of the quantum channel has an attenuation coefficient of 0.2dB/km, which refers to the transmittance of the quantum channel $\eta_{ch}$=$10^{-0.2L/10}$ with the fiber length of $L$ (km).

The error probability when Alice sends a pulse with the average photon number $\mu_a$ ($a$= S, D, V) in $x$-basis ($x$=Y, Z) is calculated with $e_{ax}$=$P_{dc}$+$e_{opt}$[1-exp(-$\eta\mu_a P_{xb}$)]+$P_{ap}D_{ax}$/2, where $e_{opt}$=0.01 is the error due to the imperfection of the optics, $\eta$ is the total detection efficiency ($\eta=\eta_{ch}\eta_{Bob}\eta_{det}$). $D_{ax}$ is the expected detection rate in *x*-basis detectors (excluding after-pulse effect) given as $D_{ax}$=1-(1-2$P_{dc}$)exp(-$\eta\mu_a P_{xb}$) for the pulse of the average photon number $\mu_a$ in *x*-basis.

We set that our QKD is $\varepsilon_{sec}$-secret and $\varepsilon_{cor}$-correct. Here $\varepsilon_{sec}$-secret means that the secret key is distinguishable from the ideal key with probability at most of $\varepsilon_{sec}$, and $\varepsilon_{cor}$-correct means that the probability of Alice and Bob sharing identical secret key is no smaller than 1-$\varepsilon_{cor}$. In the key distillation process, we assume that the error correction cost is given by $\lambda_{EC} = f_{EC} h(e_Z)$ with $f_{EC} = 1.2$ and $e_Z = (D_{SZ} e_{SZ} + D_{DZ} e_{DZ} + D_{VZ} e_{VZ})/(D_{SZ} + D_{DZ} + D_{VZ})$. We employ $\varepsilon_{sec} = 2 \times 10^{-11}$ and $\varepsilon_{cor} = 2^{-127}$ in the secure key rate simulation.

The simulated secure key rate per pulse for several valid intensity ranges from $0.2\sigma_a$ to $1.0\sigma_a$ for 100 Mbits sifted key as functions of transmission length are shown in Fig. 3. This figure implies that smaller valid range leads to longer distance. On the other hand, regarding key generation rate at short and middle distance less than 70 km, around $0.6\sigma_a$ is optimal because of the trade-off between the amount of eliminated pulses by IS and the amount of discarded bits in the privacy amplification due to the effect of intensity distribution. Note that the optimal intensity range highly depends on the characteristics of the QKD system. Therefore, to maximize secure key rate, we need careful parameter selections according to the intensity fluctuation levels in the real system.

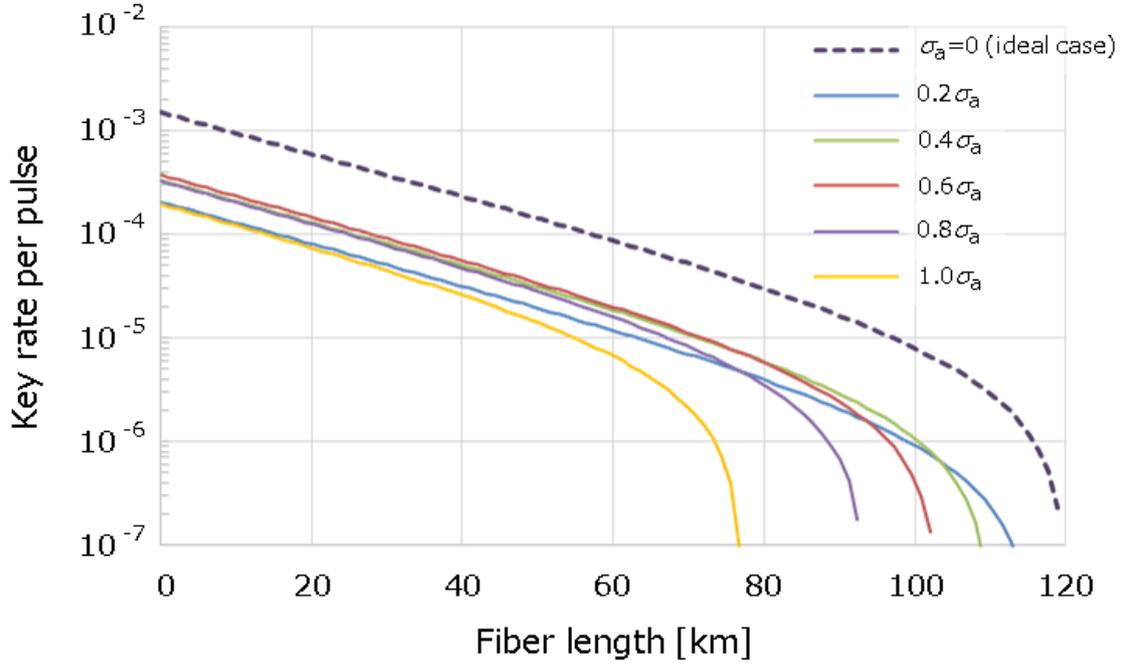

**Figure 3: Simulation of the final key rate for 100 Mbits sifted key considering intensity fluctuation caused by pattern effects and random noise.** Valid intensity range of IS is changed from $0.2\sigma_a$ to $1.0\sigma_a$ (*a*= S, D, V).

## Conclusion

We have pointed out and experimentally evaluated intensity fluctuations of each optical pulse for 1.24GHz-clocked high-speed QKD system for the first time. We found large intensity deviation of decoy pulse depending on previous modulation pattern due to distortion of electric signals originated from the limited bandwidth of the electronics. We newly developed countermeasures named "pattern sifting" and "alternate key distillation" against the correlated deviation, which aim at recovering the IID assumption common to the most of security proofs. We further showed that the remaining random intensity distribution due to thermal noise or timing jitter can be handled with "intensity sifting" method, which enables us to generate secure key with finite-length analysis using a real GHz-clock QKD system. The developed countermeasures yield reasonable key after 100-km transmission. Our results provide simple and effective solution to wide range of high-speed QKD systems, where the signal distortion is observed.

## Methods

In Methods, we will prove security of the proposed decoy-state BB84 protocol under the pattern effect. The protocol uses the pattern sifting (PS) and the alternate key distillation (AKD). These two methods enable us to attune security proofs for standard decoy-state BB84 protocols to prove our case. To represent existing analyses of standard decoy-state BB84 protocols, we summarize notations as follows.

- **a**: a sequence whose element $a_i \in \{S, D, V\}$ represents the type (Signal, Decoy, or Vacuum, respectively) of the $i$-th pulse.
- **n**: a sequence whose element $n_i \in \{0, 1, \cdots\}$ represents the number of photons emitted in the $i$-th pulse.
- $\mathbf{x}_A, \mathbf{x}_B$: sequences whose elements $x_{A,i}, x_{B,i} \in \{Y, Z\}$ represent choices of the basis for the $i$-th pulse by Alice and Bob, respectively.
- $\mathbf{b}_A$: a sequence whose element $b_{A,i} \in \{0,1\}$ represents Alice's bit value for the $i$-th pulse.
- $\boldsymbol{\Lambda}$: a sequence whose element $\Lambda_i$ represents the set of all the data associated with the $i$-th pulse except $a_i$ and $n_i$. It includes $x_{A,i}, x_{B,i}$ and $b_{A,i}$ as well as Bob's measurement outcome.
- $\{\mu_S, \mu_D, \mu_V\}$: mean photon numbers corresponding to the types $S$, $D$ and $V$.
- $q(n, \mu) := e^{-\mu} \mu^n / n!$: the probability of $n$ photons emitted in a pulse with a mean photon number $\mu$.

The assumptions used in existing analyses of standard decoy-state BB84 protocols are summarized as follows.

1. The sequence $\mathbf{a}$ is independent and identically distributed (IID) with prior probabilities $p_S, p_D$ and $p_V$.
2. Each of the sequences $\mathbf{x}_A$ and $\mathbf{x}_B$ is IID with given probabilities.
3. The sequence $\mathbf{b}_A$ is IID with probability $1/2$.
4. The probability distribution $\Pr(\mathbf{a},\mathbf{n})$ is written as $\prod_i f(a_i, n_i)$ with $f(a,n) = p_a q(n, \mu_a)$.
5. Conditioned on $\mathbf{a}, \mathbf{n}, \mathbf{x}_A$ and $\mathbf{b}_A$, the state of the whole pulses is written as $\bigotimes_i \rho(n_i, x_{A,i}, b_{A,i})$.
6. The three sets of variables $\mathbf{a}, \mathbf{n}$ and $\Lambda$ form a Markov chain, which we denote by $\mathbf{a} \to \mathbf{n} \to \Lambda$.

Assumption 6 is not an independent assumption but is a consequence of assumption 5 and the independence of $\mathbf{a}$ from $\mathbf{x}_A$ and $\mathbf{b}_A$. We have included it for convenience of discussions below.

By use of them, we represent existing security analyses of standard decoy-state BB84 protocols as a combination of two arguments (a) and (b). The argument (a) is a decoy-state analysis, which makes estimation over a photon number distribution. The decoy-state analysis is purely mathematical and the only assumptions it uses are 4 and 6. The result of estimation is usually given as a set of inequalities that are satisfied except with a small probability $\varepsilon_a$. The inequalities imply, for example, a lower bound on the number of detections from single-photon ($n=1$) signals. For our purpose, it is convenient to represent these inequalities equivalently by using a set $\Gamma$ of admissible values $(\mathbf{a},\mathbf{n},\Lambda)$, namely, as $(\mathbf{a},\mathbf{n},\Lambda) \in \Gamma$.

The argument (b) is a BB84 analysis with a known photon number distribution. It provides a rule $l(\mathbf{a},\Lambda)$ to determine the length $l$ of the final key from the data available in the protocol, and proves that it is secure if $(\mathbf{a},\mathbf{n},\Lambda) \in \Gamma$ holds. We emphasize here that this part of the argument does not rely on assumption 4 any longer since it only cares about the security in the case of $(\mathbf{a},\mathbf{n},\Lambda) \in \Gamma$. To describe the argument (b) more precisely, let us describe the real protocol as a diagram given in Fig. 4, in which the box "key substitution" should be ignored. We also introduce the ideal protocol, in which the actual key is substituted by an ideal key. The real protocol is called $\varepsilon$-secure if it is distinguishable from the ideal protocol by at most $\varepsilon$, measured in terms of trace distance.

The argument (b) does not prove the security of the real protocol, but that of a variant which we call the intermediate protocol. In the intermediate protocol, the actual key is substituted by an ideal key if and only if $(\mathbf{a},\mathbf{n},\Lambda) \notin \Gamma$ holds. The fact that the argument (b) does not rely on assumption 4 implies that the security is not threatened even if $\mathbf{a}$ and $\mathbf{n}$ are determined by an adversary. Let us call the shaded region in Fig. 4 as the sub-protocol, which regards $\mathbf{a}$ and $\mathbf{n}$ as the data provided from outside. What is actually proved in the argument (b) is the security of the intermediate sub-protocol, or its indistinguishability from the ideal sub-protocol. The statements of arguments (a) and

(b) are summarized as follows.

    (a) For a positive real number $\varepsilon_a$ and a set $\Gamma$, $\Pr((\mathbf{a},\mathbf{n},\Lambda) \notin \Gamma) < \varepsilon_a$.

    (b) The intermediate sub-protocol with a set $\Gamma$ and the final key length $l(\mathbf{a},\Lambda)$ is $\varepsilon_b$-secure for a positive real number $\varepsilon_b$.

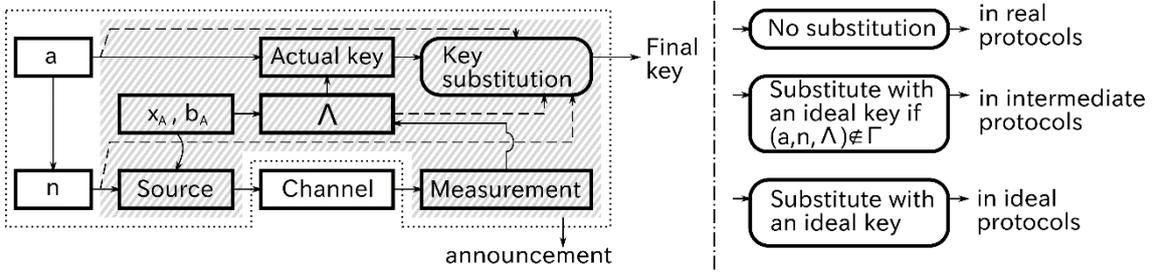

**Figure 4: A schematic representing a standard decoy-state BB84 protocol.** Depending on real, intermediate, and ideal protocols, the rounded-corner box works differently as written in the right side of this figure. The shaded area represents the sub-protocol.

    Argument (b) guarantees that the intermediate sub-protocol is $\varepsilon_b$-secure for any $\mathbf{a}$ and $\mathbf{n}$. Hence, the intermediate protocol, which uses the actual $\mathbf{a}$ and $\mathbf{n}$ as an input of the sub-protocol, is also $\varepsilon_b$-secure. Since the difference between the real protocol and the intermediate protocol arises only if $(\mathbf{a},\mathbf{n},\Lambda) \notin \Gamma$, assumption (a) implies that the trace distance is no larger than $\varepsilon_a$. Using the triangle inequality for the trace distance, the real protocol is proved to be $(\varepsilon_a + \varepsilon_b)$-secure.

    Now, we consider a protocol under the pattern effect, which means the $i$-th type $a_i$ affects $n_{i+1}$ as well as $n_i$. We assume a model in which the mean photon number of the $i$-th pulse is represented as $\mu(a_i, a_{i-1})$ which satisfies

$$\mu(S, a_{i-1}) = \mu_S, \ \mu(D, S) = \mu_D, \ \mu(V, a_{i-1}) = \mu_V \tag{1}$$

for any $a_{i-1}$. The probability distribution of $(\mathbf{a},\mathbf{n})$ is then written as

$$\Pr(\mathbf{a},\mathbf{n}) = \prod_i \tilde{f}(a_i, a_{i-1}, n_i), \tag{2}$$

where $\tilde{f}(a_i, a_{i-1}, n_i) = p_{a_i} q(n_i, \mu(a_i, a_{i-1}))$. Although this change threatens assumption 4 in the standard case, we will show that the analyses in the standard case can be applied to the elements after PS and AKD, such as the even-indexed and pattern-sifted elements. To represent the restriction on the even-indexed elements, the odd-indexed elements and the pattern-sifted elements, we use the superscripts "even", "odd" and "PS", such as $\mathbf{a}^{\text{even}}, \mathbf{n}^{\text{odd,PS}}$ and so on.

    We define a set of even indices as $I^{\text{even}}$ and a set of indices of the even-indexed and pattern-sifted elements as

$$I^{\text{even,PS}} := \{i \mid i \in I^{\text{even}}, a_{i-1} = S, a_{i+1} \in \{S,V\}\}. \tag{3}$$

It is then easy to see that $\tilde{f}(a_i, a_{i-1}, n_i) = f(a_i, n_i)$ and $\tilde{f}(a_{i+1}, a_i, n_{i+1}) = f(a_{i+1}, n_{i+1})$ hold for $i \in I^{\text{even,PS}}$.

Although the pattern effect disturbs the form of $\Pr(\mathbf{a}, \mathbf{n})$ and prevents us from directly applying (a), we will show that the even-indexed and pattern-sifted elements satisfies assumptions 4 and 6 conditionally on $(\mathbf{a}^{\text{odd}}, \mathbf{n}^{\text{odd}})$, namely, the following two properties hold:

(i) $\Pr(\mathbf{a}^{\text{even,PS}}, \mathbf{n}^{\text{even,PS}} | \mathbf{a}^{\text{odd}}, \mathbf{n}^{\text{odd}}) = \prod_{i \in I^{\text{even,PS}}} f(a_i, n_i).$ (4)

(ii) $\mathbf{a}^{\text{even,PS}} \to (\mathbf{n}^{\text{even,PS}}, \mathbf{n}^{\text{odd}}, \mathbf{a}^{\text{odd}}) \to \Lambda^{\text{even,PS}}.$ (5)

To show (i), we focus on the fact that the sequence of the pairs $(a_i, n_i)$ forms a Markov chain $(a_1, n_1) \to (a_2, n_2) \to (a_3, n_3) \to \cdots$ under the pattern effect. It means

$$\Pr(\mathbf{a}^{\text{even}}, \mathbf{n}^{\text{even}} | \mathbf{a}^{\text{odd}}, \mathbf{n}^{\text{odd}}) = \prod_{i \in I^{\text{even}}} \Pr(a_i, n_i | a_{i-1}, n_{i-1}, a_{i+1}, n_{i+1}).$$ (6)

For $i \in I^{\text{even,PS}}$, we find

$$\Pr(a_i, n_i | a_{i-1}, n_{i-1}, a_{i+1}, n_{i+1}) = \frac{\tilde{f}(a_{i+1}, a_i, n_{i+1}) \tilde{f}(a_i, a_{i-1}, n_i)}{\sum_{a_i', n_i'} \tilde{f}(a_{i+1}, a_i', n_{i+1}) \tilde{f}(a_i', a_{i-1}, n_i')}$$

$$= \frac{f(a_{i+1}, n_{i+1}) f(a_i, n_i)}{\sum_{a_i', n_i'} f(a_{i+1}, n_{i+1}) f(a_i', n_i')}$$

$$= f(a_i, n_i),$$ (7)

and it means that the property (i) holds.

To show (ii), we remind that the Markov property $\mathbf{a} \to \mathbf{n} \to \Lambda$ holds even under the pattern effect. It means $\mathbf{a}^{\text{even}} \to (\mathbf{n}^{\text{even}}, \mathbf{n}^{\text{odd}}, \mathbf{a}^{\text{odd}}) \to \Lambda$, and by restricting $\mathbf{a}$ and $\Lambda$, we obtain $\mathbf{a}^{\text{even,PS}} \to (\mathbf{n}^{\text{even}}, \mathbf{n}^{\text{odd}}, \mathbf{a}^{\text{odd}}) \to \Lambda^{\text{even,PS}}$. Eq. (6) means that the even-indexed pairs $\{(a_{2j}, n_{2j})\}_j$ become independent from each other if we fix $(\mathbf{a}^{\text{odd}}, \mathbf{n}^{\text{odd}})$. It leads to $\mathbf{a}^{\text{even,PS}} \to (\mathbf{n}^{\text{even,PS}}, \mathbf{n}^{\text{odd}}, \mathbf{a}^{\text{odd}}) \to \mathbf{n}^{\text{even},\overline{\text{PS}}}$, where the superscript $\overline{\text{PS}}$ means the elements removed by the pattern sifting. In general, two Markov properties $X \to Y_1 \to Y_2$ and $X \to (Y_1, Y_2) \to Z$ mean $X \to Y_1 \to (Z, Y_2)$, leading to $X \to Y_1 \to Z$. Setting $(X, Y_1, Y_2, Z)$ to be $(\mathbf{a}^{\text{even,PS}}, (\mathbf{n}^{\text{even,PS}} \mathbf{n}^{\text{odd}}, \mathbf{a}^{\text{odd}}), \mathbf{n}^{\text{even},\overline{\text{PS}}}, \Lambda^{\text{even,PS}})$, we obtain the condition (ii).

Since both conditions (i) and (ii) required for (a) are satisfied in the even-indexed and pattern-sifted elements, we can apply (a) to them and obtain

$$\Pr((\mathbf{a}^{\text{even,PS}}, \mathbf{n}^{\text{even,PS}}, \Lambda^{\text{even,PS}}) \notin \Gamma | \mathbf{a}^{\text{odd}}, \mathbf{n}^{\text{odd}}) < \varepsilon_a.$$ (8)

It also means

$$\Pr((\mathbf{a}^{even,PS}, \mathbf{n}^{even,PS}, \mathbf{\Lambda}^{even,PS}) \notin \Gamma) < \varepsilon_a. \qquad (9)$$

because $\varepsilon_a$ does not depend on $\mathbf{a}^{odd}$ and $\mathbf{n}^{odd}$. The same goes for the odd-indexed and pattern-sifted elements, and we can use the union bound to obtain

$$\Pr((\mathbf{a}^{even,PS}, \mathbf{n}^{even,PS}, \mathbf{\Lambda}^{even,PS}) \notin \Gamma \vee (\mathbf{a}^{odd,PS}, \mathbf{n}^{odd,PS}, \mathbf{\Lambda}^{odd,PS}) \notin \Gamma) < 2\varepsilon_a. \qquad (10)$$

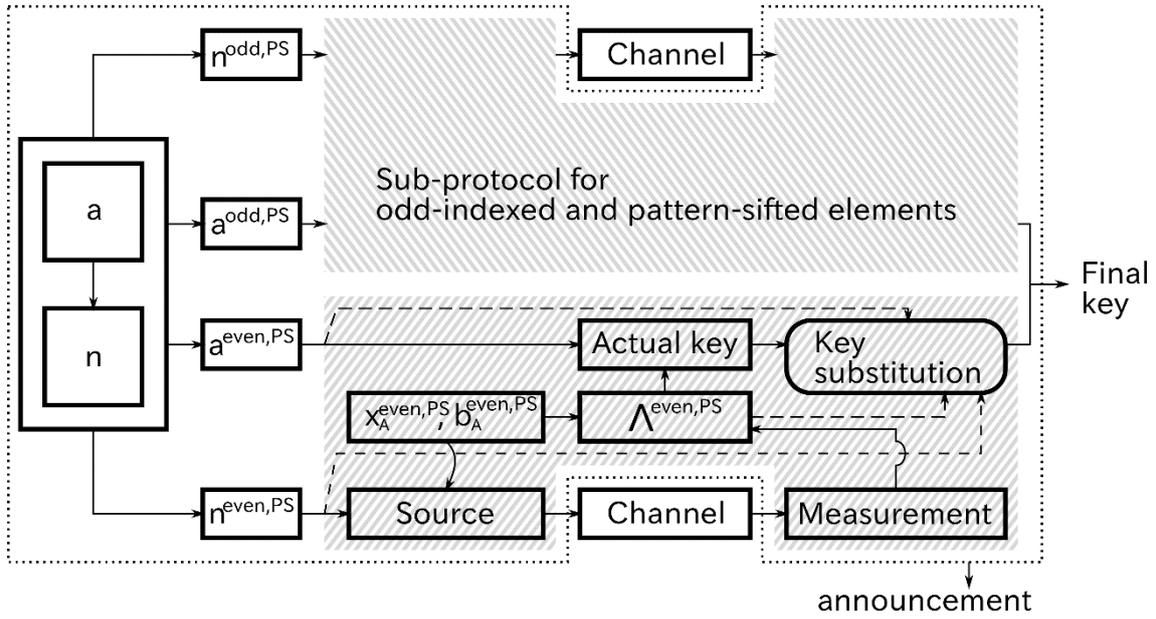

**Figure 5: A schematic representation of a decoy-state BB84 protocol with PS and AKD.** The protocol includes two sub-protocols defined in Fig. 4. Depending on real protocol, intermediate and ideal protocols, the sub-protocols change as in Fig. 4.

Next, we consider the whole protocol with PS and AKD, which can be regarded as follows (see Fig. 5). The protocol generates $\mathbf{a}$ and $\mathbf{n}$, generates $(\mathbf{a}^{even,PS}, \mathbf{n}^{even,PS})$ and $(\mathbf{a}^{odd,PS}, \mathbf{n}^{odd,PS})$, and supplies them to two sub-protocols which are identical as the sub-protocol in Fig. 4. Each of the sub-protocols produces a final key, and the concatenation of the two keys is the output of the protocol. We define the intermediate protocols and the ideal protocols as in the standard case. The argument (b) guarantees that each of the intermediate sub-protocols is $\varepsilon_b$-secure. The standard argument of the universal composability means that the intermediate protocol is $2\varepsilon_b$-secure. Since the difference between the real protocol and the intermediate protocol is caused by the event where the condition $\left(\mathbf{a}^{even,PS}, \mathbf{n}^{even,PS}, \mathbf{\Lambda}^{even,PS}\right) \in \Gamma$ or $\left(\mathbf{a}^{odd,PS}, \mathbf{n}^{odd,PS}, \mathbf{\Lambda}^{odd,PS}\right) \in \Gamma$ is not satisfied, Eq. (10) bounds the trace distance to be no larger than $2\varepsilon_a$. As a consequence, we find that the real protocol is

$2(\varepsilon_a + \varepsilon_b)$-secure.

The above proof is also applicable when there exist independent fluctuations in the mean photon number after PS and AKD in our experiment. It can be done by extending a function $f(a,n)$ to a set of functions satisfying a condition about intensity fluctuations. This change does not affect the above reasoning as long as the choice of set $\Gamma$ in the argument (a) is dictated from a proof accommodating such fluctuations.

This work was supported in part by the ImPACT Program of the Cabinet Office Japan.

**References**

1. Bennett, C. H. & Brassard, G. Quantum cryptography: Public-key distribution and coin tossing, *Proceedings IEEE Int. Conf. on Computers, Systems and Signal Processing*, Bangalore, India, pp. 175–179 (IEEE, New York, 1984).
2. Gisin, N., Ribordy, G., Tittel, W. & Zbinden, H. Quantum cryptography. Rev. Mod. Phys., **74**, 145–195 (2002).
3. Scarani, V., Bechmann-Pasquinucci, H., Cerf, N. J., Dušek, M., Lütkenhaus, N., & Peev, M., The security of practical quantum key distribution. Rev. Mod. Phys. **81**, 1301–1350 (2009).
4. Fujiwara, M., Waseda, A., Nojima, R., Moriai, S., Ogata, W., & Sasaki, M. Unbreakable distributed storage with quantum key distribution network and password-authenticated secret sharing. Sci. Reports, **6**, 28988-1-8 (2016).
5. Braun, J., Buchmann, J., Demirel, D., Geihs, M., Fujiwara, M., Moriai, S., Sasaki, M., & Waseda, A. LINCOS - A Storage System Providing Long-Term Integrity, Authenticity, and Confidentiality. *Proc. the 2017 ACM on Asia Conference on Computer and Communications Security*, 461-468 (2017).
6. Elliott, C., Colvin, A., Pearson, D., Pikalo, O., Schlafer, J., & Yeh, H. Current status of the DARPA quantum network. *Proc. SPIE* 5815, 138–149 (2005); arXiv:quant-ph/0503058v2.
7. Stucki, D. et al. Long-term performance of the SwissQuamtum quantum key distribution network in a field environment. New J. Phys. 13(12), 123001, 1-18 (2011).
8. Peev, M. et al, The SECOQC quantum key distribution network in Vienna. New J. Phys. **11**(7), 075001/1-37 (2009).
9. Sasaki, M. et al. Field test of quantum key distribution in the Tokyo QKD Network. Opt. Express, **19**(11), 10387–10409 (2011).
10. Wang, S., et al. Field and long-term demonstration of a wide area quantum key distribution network. Opt. Express **22**(18), 21739-21756 (2014).
11. http://spectrum.ieee.org/telecom/security/chinas-2000km-quantum-link-is-almost-complete. Confirmed 17th May (2017).
12. Sasaki, M. et al. Quantum Photonic Network: Concept, Basic Tools, and Future Issues. J. Selected

Topics in Quant. Elec., 21(3), 6400313 (2015).

13. Tanaka, A., et al., High-speed quantum key distribution system for 1-Mbps real-time key generation. IEEE trans. Quant. Elec. 48(4), 542-550 (2012).

14. Dynes, J. F. et al. Stability of high bit rate quantum key distribution on installed fiber. Opt. Express **20**, 16339-16347 (2012).

15. Yoshino, K., Ochi, T., Fujiwara, M., Sasaki, M. & Tajima, A. Maintenance-free operation of WDM quantum key distribution system through a field fiber over 30 days. Opt. Express **21**, 31395-31401 (2013).

16. Zhao, Y., Fung, C.-H.F., Qi, B., Chen, C., & Lo, H.-K. Quantum hacking: Experimental demonstration of time-shift attack against practical quantum-key-distribution systems. Phys. Rev. A **78**, 042333 (2008).

17. Lydersen, L., Wiechers, C., Wittmann, C., Elser, D., Skaar, J., & Makarov, V. Hacking commercial quantum cryptography systems by tailored bright illumination. Nat. Photonics **4**, 686-689 (2010).

18. Yuan, Z., L., Dynes, F., J., & Shields, A., J. Avoiding the detector blinding attack on quantum cryptography. Nat. Photonics **4**, 800-801 (2010).

19. Jain, N., Stiller, B., Khan, I., Makarov, V., Marquardt, C., & Leuchs, G. Risk analysis of Trojan-horse attacks on practical quantum key distributions. IEEE J. Selected Topics in Quamt. Elec. **21**(3) 6600710 (2015).

20. Lo, H.-K., Curty, M. & Qi, B. Measurement-device-independent quantum key distribution. Phys. Rev. Lett. 108, 130503 (2012).

21. Rubenok, A., Slater, J. A., Chan, P., Lucio-Martinez, I. & Tittel, W. Real-world two-photon interference and proof-of-principle quantum key distribution immune to detector attacks. Phys. Rev. Lett. 111, 130501 (2013).

22. Kobayashi, K., Tomita, A., & Okamoto, A. Evaluation of the phase randomness of a light source in quantum-key-distribution systems with an attenuated laser. Phys. Rev. A**90**, 032320 (2014).

23. Tamaki, K., Curty, M., Kato, G., Lo, H.-K., & Azuma, K., Loss-tolerant quantum cryptography with imperfect sources. Phys. Rev. A **90**, 052314 (2014).

24. Lucamarini, M., Choi, I., Ward, M., B., Dynes, J., F., Yuan, Z., L., & Shields, A., J. Practical security bounds against the Trejan-Hose attack in quantum key distribution. Phys. Rev. X**5**, 031030-1-19 (2015).

25. Tamaki, K., Curty, M., & Lucamarini, M., Decoy-state quantum key distribution with a leaky source. New J. Phys. 5, 065008 (2016).

26. Nakata, K., Tomita, A., Fujiwara, M., Yoshino, K., Tajima, A., Okamoto, A., & Ogawa, K., Intensity fluctuation of a gain-switched semiconductor laser for quantum key distribution systems. Opt. Express **25**, 622 -634 (2017).

27. Nagamatsu, Y., Mizutani, A., Ikuta, R., Tamamoto, T., Imoto, N., & Tamaki, K. Security of

quantum key distribution with light sources that are not independently and identically distributed. Phys. Rev. A **93** (4), 042325-1-10 (2016).

28. Wang, Y., Bao, W. S., Zhou, C., Jiang, M. S., & Li, H. W. Tight finite-key analysis of a practical decoy-state quantum key distribution with unstable sources. Phys. Rev. A **94**(3), 032335 (2016).

29. Zhang, Z., Zhao, Q., Razavi, M., & Ma, X. Improved key-rate bounds for practical decoy-state quantum-key-distribution sytems. Phys. Rev. A **95**, 012333-1-14 (2017).

30. Lo, H.-K., Ma, X. & Chen, K. Decoy state quantum key distribution. Phys. Rev. Lett. **94**, 230504 (2005).

31. Ma, X., Qi, B., Zhao, Y. & Lo, H.-K. Practical decoy state for quantum key distribution. Phys. Rev. A **72**, 012326 (2005).

32. Lim, C. C. W., Curty, M., Walenta, N., Xu, F., & Zbinden, H. Concise security bounds for practical decoy-state quantum key distribution. Phys. Rev. A **89**(2), 022307 (2014).